\documentclass[11pt]{article}

\usepackage{setspace,graphicx,epstopdf,amsmath,amsfonts,amssymb,amsthm,versionPO}
\usepackage{marginnote,datetime,enumitem,subfigure,rotating,fancyvrb}
\usepackage{hyperref,float}
\usepackage[utf8]{inputenc}
\usepackage[longnamesfirst]{natbib}
\usdate

\excludeversion{notes}		% Include notes?
\includeversion{links}          % Turn hyperlinks on?

\iflinks{}{\hypersetup{draft=true}}

\ifnotes{%
\usepackage[margin=1in,paperwidth=10in,right=2.5in]{geometry}%
\usepackage[textwidth=1.4in,shadow,colorinlistoftodos]{todonotes}%
}{%
\usepackage[margin=1in]{geometry}%
\usepackage[disable]{todonotes}%
}

\makeatletter\let\chapter\@undefined\makeatother % Undefine \chapter for todonotes

\usepackage{indentfirst} % Indent first sentence of a new section.
\usepackage{endnotes}    % Use endnotes instead of footnotes
\usepackage{jf}          % JF-specific formatting of sections, etc.
\usepackage[labelfont=bf,labelsep=period]{caption}   % Format figure captions
\begin{document}

\setlist{noitemsep}  % Reduce space between list items (itemize, enumerate, etc.)
\onehalfspacing      % Use 1.5 spacing
% Use endnotes instead of footnotes - redefine \footnote command
\renewcommand{\footnote}{\endnote}  % Endnotes instead of footnotes

\author{TIAGO FERNANDES\thanks{\rm Tiago is with the Department of Applied Physics, Institute of Physics, University of São Paulo, Brazil.
Acknowledgments to Francisco E. M. da Silveira and Lissa Campos for reading the manuscript, corrections and suggestions, and 
to \emph{Brazilian Coordination for the Improvement of Higher Education Personnel} (CAPES-BR), for my PhD scholarship.
Contact info: ORCiD: 0000-0001-8089-0506, tiago2.fernandes@usp.br.}}

\title{\Large \bf Some Physics Notions on Monetary Standard}

\date{}              % No date for final submission

% Create title page with no page number

\maketitle
\thispagestyle{empty}

\bigskip

\centerline{\bf ABSTRACT}

\begin{doublespace}  % Double-space the abstract and don't indent it
  \noindent Regardless of the gold-standard being considered as outdated, it provides valuable signs concerning the development of novel 
monetary standards, better adjusted to the current macroeconomic environment. By using a point of view of classical physics, the intent of 
this work is doing a review of the concept of monetary standard and show that the energy matrix of an economy together with a new monetary 
standard, based on the energy supply capacity, can play an essential role in the sustainable growth.

\end{doublespace}

%\keywords{Monetary Standard, Physics, Energy Supply}

\medskip

\vfill
\noindent\textbf{Disclosure statement:} I have nothing to disclose.

\noindent\textbf{Conformity declaration:} I have no conflict of interests on the subject treated in this work.
\vfill

\noindent\textbf{Revision history:}
\begin{Verbatim}[fontsize=\scriptsize]
V0.1 - initial version;
V0.2 - inclusion of Standard-Capacity equivalence and cosmetics;
V0.3 - cosmetics and references update;
V0.4(this version) - literature review, inclusion of ``Related works'', ``Economic activity and inflation'',
                     Nash ``last problem'' and ... cosmetics. :-)
\end{Verbatim}
\clearpage

\section{Introduction} \label{Introduction}

Concepts and mathematical methods of many sub-fields of physics such as thermodynamics, statistical mechanics, complex systems, chaos 
theory and many others, are often imported and employed in the economic sciences (\citealt{ECONOFISICA2}). Supported by a growing number of 
studies that take the use of these techniques, this trend has been relatively successful in describing macro and microeconomics behavior and 
becoming an important research branch in economics, as pointed out by (\citealt{ECONOFISICA1}).

The knowledge advances concerning finance and economic dynamics, provided by this branch, show up able to produce and sediment a 
significant conceptual change in the understanding of the economy. Nevertheless, it is natural a self-questioning about how the methods and 
concepts of the natural sciences can be to contribute to expanding the knowledge of the observed economic phenomena, whereas econophysics is 
still a recent branch and has not been yet completely accepted (\citealt{ECONOFISICA0, ECONOFISICA6}), despite the important results 
reached in several works.

In this context, this work is an attempt to apply the physics concepts to the economic thinking. Meanwhile, we employ a different approach 
from the econophysics. Using a critical analysis, we identify the physical notions linked to the monetary standard and also show the 
possible conflicts among the associated physical and economic concepts. Taking the gold-standard as the initial model and coupled with our 
analyzes, we also propose a modified form of commodity monetary standard, now based on energy supply capacity (ESC). Although the 
gold-standard is an outdated concept, it provides a secure basis for the development of novel monetary forms, as developed here. Thus, we 
encourage the opening of discussion about the consequences of the model adoption by an economy, mainly on the demand for renewable energy 
sources.

\section{Gold-Standard in a Physical Perspective} \label{matter_standard}

The functions of money are, in general, dependents of the attributed value notion. If a currency has no value, then, cannot be fulfill the 
monetary function. For reasons that we will not explore here, money standardization (when employed) has the predicate of the converting 
money in exchange for a certain amount of precious metals and accordingly, the substance has an intrinsic monetary 
value (\citealt{GOLDSTANDART}).

Note that, from the physical and chemical point of view, the noble metals normally used to monetary standardization do not have any special 
characteristics that can distinguish them in terms of value to other chemical elements listed in the periodic table. Propitious 
technological factors are important for obtaining and purification of gold and silver, since that, among rare metals, they are the most 
easily obtained. However, the acknowledged value of these metals is due purely by tradition and psychological questions propagated through 
the successive generations (\citealt{PSYCHOLOGY}). Thereby, the option by gold is an important and widely accepted convention.

Regardless of these conventions, we can identify an interrelationship between the concept of the gold standard with the physical concept of 
matter (quantity) and we let's label it by \emph{matter-standard}. However, even in physics, the matter is an abstract concept with timely 
definitions, depending on the context under study (classical mechanics, relativity, particle and fields physics, and others). In this, the 
monetary standard indicates a \emph{capacity} to supply the quantity of a specified type of chemical element, to which is being the holder 
of the intrinsic monetary value.

The essential physical characteristics, directly associated with the matter standard, are below listed:
\begin{enumerate}
 \item \textbf{{Invariance:}} Refers to the behavior of matter about it is space-time stability. In the matter-standard, this feature is 
important because otherwise, it will be difficult to convince people that a material element has a comparative value if it changes over 
time 
or is differ depending on the position.
  \item \textbf{{Abundance:}} Refers to the available and recoverable quantity of the material in nature. Is required that must be abundant 
enough to represent all the circulating money and rare at the same time to not have trivialized value.
  \item \textbf{{Capacity notion:}} The amount of money represented is directly proportional to the amount of matter available. 
Consequently, the matter standard is directly related to a notion of supply capacity.
  \item \textbf{{Comparability:}} Some physical feature of the material should allow comparisons between the samples. Is notable that the 
amount of matter is a comparable feature.
\end{enumerate}

The space position also conveys the notion of security: the matter-standard is present in a specific place (reserve banks) and thus, it 
provides a conversion confidence; but does not represent an intrinsic physical characteristic of the material. The item 4 could be naively 
interpreted as a monetary version of the zero law of thermodynamics. Is necessary to emphasize that the notion of comparability of matter 
quantity encapsulates the notion of value, attributed by people even to the base element of matter-standard. Each person may have it is 
a own version of the zero law and thus, a thermodynamic analogous will be subject to billions of Maxwell demons (\citealt{MAXWELL_DEMON}).

Applying these notions, we conclude that various chemical elements can close with the previous listing of characteristics to a matter-based 
standardization, wherefore, even restricted, the list does not to make the gold element a single class representative. Strictly from the 
physical point of view, any monetary form based on matter quantity is equivalent to any other. Perception of value is human inherently.

\section{Non-matter standard - Energy Based Monetary Standard} \label{Non_Matter_Standard}

In this section, we will explore the possible entities and physical concepts that are candidates for a monetary standard. From the 
standpoint of classical physics, the known universe may be divided into four constituents: \emph{matter, energy, space and time}. We point 
out that, accordingly the modern physical conception shows the equivalence among some of these entities, but we are still seeking a theory 
capable of agglutinating all these entities within the same formalism (\citealt{TEORIA_UNIFICADA}). Situations as this, or even in cases 
where the use of special or general relativity is necessary, are relevant to scientific experiments in controlled laboratory environments, 
astrophysics, high-energy physics or cosmology (\citealt{APLIC_RELATIVIDADE}). Conversely, they are not significant in the daily markets 
and the financial environment.

Moreover, space and time are not susceptible to manipulation by humans in the current technological context and it is also still quite 
controversial whether they may be manipulated in some future (\citealt{TIMETRAVEL01, VIAGENS_TEMPO, TIMETRAVEL02}). Consequently, just the 
energy concept remains to be explored as a monetary element. Energy is also an abstract entity and cannot be directly observed, 
nevertheless, this does not prevent us from analyzing some characteristics that may make it an interesting candidate to compose a monetary 
standard and how it may be implemented in practice. The characteristics of interest in the concrete are listed below:
\begin{enumerate}
 \item \textbf{{Storable and convertible:}} The characteristic of the energy of being able to be accumulated in some volume as potential 
energy and converted into other forms of energy. These techniques are largely dominated by the current technological maturity, for example 
in hydroelectric power plants.
 \item \textbf{{Usable to perform work:}} The work quoted here is in the physical sense, but it is a more relevant characteristic in the 
economic and financial sense, since through it are processed the goods and services.
 \item \textbf{{Conservation:}} Naturally, it is implied that the system is closed. This characteristic is a consequence of the invariance 
of the physics laws by temporal translations, as stated by (\citealt{Noether1918}). In open systems, a portion of energy flows out of the 
system domains\footnotemark. \footnotetext[1]{ A more applied description to the Noethers theorem is present in (\citealt{NOETHER}).}
\end{enumerate}

Note that these are characteristics of energy and not directly of an energy-based monetary standard. However, the energy conservation is 
identified with the invariance characteristic of matter-standard and storage with the notion of supply capacity, per function analogy. About 
to comparability, this function may be accomplished by the ability to perform work, whose comparison means may be fulfilled by the existing 
international standard of measurement. Abundance is a bit more complex seeing that it depends on the kind of energy used, since it is 
usually converted from a repository to a more easily usable form of energy, such as the electric energy converted from level and flow in a 
water dam. But the conversion is always done on demand. Thus, all the characteristics of a matter standard may also be achieved by some 
energy feature.

Compared to the matter-standard case, both have finite abundance. In addition, depending on the energy matrix of the economy, the ESC is 
potentially small regarding the time evolution of demand. Thus, renewable or long-term power sources may insure the sustainable monetary 
growth of the markets. We emphasize that, the common point in both cases, matter and energy, that is the central physical characteristic 
that endorses the use as monetary-standard is linked with the supply capacity. This form of standardization was previously identified in 
section \ref{Introduction} by the term \emph{``modified commodity standard''} because indeed, the standard is based on the intrinsic value 
of the supply capacity and not on the intrinsic value of the product itself.
\\

\noindent\textbf{Standard-Capacity equivalence:} The monetary consolidation of the ESC may be accomplished in several ways, but here we 
propose a simplified one's, just for the establishing the procedure more explicit. For example: On the commercial scale, the electric 
energy is usually measured in $kWh$ but due the scale of production capacity of a country is substantially higher, we should consider the 
$GWy$ (gigawatt-year) or $TWy$ (Terawatt-year) as more appropriate unit and thus, the conversion to an arbitrary monetary unit in a purely 
attached system (all-gold like) is given by:
\begin{equation}
M = \sum\limits_n {{c_n}{{\left\langle {{\varepsilon _n}} \right\rangle }_{\Delta T_{n}}}}
\label{eq:Conversion_E_to_M}
\end{equation}
, where: $M$ is the total represented money in an arbitrary monetary unit $\left[ {a.m.u.} \right]$, ${{\varepsilon _n}}$ is total energy 
supply capacity in $\left[ {GWy} \right]$ of nth source, $c_n$ is a conversion factor in $\left[ {a.m.u.} \right]{\left[ {GWy} \right]^{ 
- 1}}$. The average operator ${\left\langle \cdot \right\rangle _{\Delta T}}$ was used to properly account the seasonalities of the energy 
sources over a suitable time interval. The time intervals choice is not necessarily the same for all energy sources. It must depend on the 
sources peculiarities. Nevertheless, there does not a mandatory presumption in favor of the equation \ref{eq:Conversion_E_to_M} and so, 
other versions are possible. See, for example, that the constant $c_n$ is individual for each source and thus, different types can be merged 
(except that the coefficients will not have the same dimensional unit). The basis of equation \ref{eq:Conversion_E_to_M} is very simple, it 
avoids money held on the banking system and neglect other kinds of hold/source agent. The conceptual clarity of the standard, on the other 
hand, is kept intact.

Furthermore, the total abundance $(A_{Total})$ can be expressed by:
\begin{equation}
A_{Total} = \sum\limits_n {{{\left\langle {{\varepsilon _n}} \right\rangle }_{\Delta T_{n}}}}
\label{eq:Abundance}
\end{equation}

The expressions \ref{eq:Conversion_E_to_M} and \ref{eq:Abundance} show that: for a given $m$, if $c_n$ is time independent and
$\frac{d}{{dt}}{\left\langle {{\varepsilon _m}} \right\rangle _{year}} < 0$ then, it is necessary that $\sum\limits_{s \ne m} 
{c_{s}\frac{d}{{dt}}{{\left\langle {{\varepsilon _s}} \right\rangle }_{year}}}  >  - c_{m}\frac{d}{{dt}}{\left\langle {{\varepsilon _m}} 
\right\rangle _{year}}$, to insure monetary growth. In simple terms, if one or more sources have their supply capabilities reducing over 
time, then the other sources need to compensate it. For other versions of equation \ref{eq:Conversion_E_to_M}, another growth conditions 
may be derived.

Note that $\epsilon_n$ is a problem to be solved by physicists and engineers, and the determination of the coefficient $c_n$ is an issue 
addressed to economists. The multidisciplinarity of this approach becomes more manifest.

\section{General Remarks}\label{general_remarks}

Nothing has been said about non-standardized currencies (fiat moneys and cryptocurrencies) because we understand the non-applicability of 
the concepts explored here. In the specific to bitcoin cryptocurrency, the literature reports a significant energy inefficiency, as 
explored in (\citealt{BITCOIN01}). But it seems clear that the value assignment from one's kind of currency to another (standardized or 
not) owes to historical robustness reasons. However, assigning an absolute value to some feature of the matter-based standard element is 
linked to an abstract concept of supply capacity. While gold has a millenary attributed and accepted monetary value, fiat moneys are more 
recent types and it is unclear an idea of supply capacity assigned to it. The monetary value of the energy-standard may be attributed in an 
analogous way from the matter, that is, the money is printed only on the basis of the installed capacity of production (energy generation 
or conversion).
\\

\noindent\textbf{Related works:} The idea of using energy as a currency is not new. The first mention was made by 
(\citealt{PROP_ENERGIA_ANTERIOR1}). This text is important, since it presents a vision adjusted to the market vocabulary. But, the energy 
is directly treated as currency and this seems not be perfectly appropriate since the capacity characteristic used was implicitly 
referenced to the ability to perform work (subject to losses, since the system is not closed). More recently the issue was addressed in 
(\citealt{PROP_ENERGIA_ANTERIOR3}). This is an interesting work that consists of an \emph{energy supply forecast integrated over a period of 
time}. Such proposal can be seen as a highly elaborate monetary rule based on energy consumption. In a subsequent work, the same author 
presents a proposal closer to a monetary standard as explored here. The difference is that he considers within the system the 
\emph{productive capacity of the embedded energy} as well \emph{energy resource potential}\footnotemark \footnotetext{Here one must be 
careful about the difference of connotation of the term ``potential'' in the physical context.} \emph{available for future expansion} 
(\citealp[p.7]{PROP_ENERGIA_ANTERIOR2}), that is, energy sources that are not yet operationally available on the matrix. Both proposals are 
conceptually different with respect to ESC. Here our concern is related to the physical and economic background interrelationship thus, in 
this sense, the works are complementary ones.

\noindent\textbf{Economic activity and inflation:} Other works have shown that the increase in energy consumption is linked to the economic 
activity growth (\citealt{ENERGYCONSUMIPTION01, ENERGYCONSUMIPTION02}). This tends to increase the investments in energy generation, as 
already pointed out in (\citealt{PROP_ENERGIA_ANTERIOR2}). It is clearly stated, with these additional considerations along with the growth 
condition of ESC-standard, the importance of the energy matrix to sustainable economic growth.

Furthermore, due to the link between the growth of the energy demand and investment in generation, the emission of money only occur on 
necessities, impacting on inflationary dynamics. There seems to be a reverse action mechanism: inflation tends to slow down the economy and 
reduce the demand for energy, thus deflating the currency. The inverse also seems true, but the changes in supply capacity or consumption 
efficiency are relatively slow. The exact mechanism of the inflation dynamics in energy supply standardized money may be further studied by 
expert researchers.

In addition, another topic that deserves further analysis is the implications of the standard on the money circulation models, as made in 
(\citealt{MONEYMODEL02, MONEYMODEL01}) and (\citealt{Yakovenko2016}) for the current non-stand money.

\noindent\textbf{Possible criticisms:} Some points of the model may be criticized, mainly on the non-representative- ness of other riches, 
since it only uses the ESC as a metric. We should note that a monetary standard does not need to represent all wealth, but it must have 
means of representing all the circulating money, provide a well-established reference and should be useful for other economic measurements.

In this context, the listing of standard characteristics, presented in the section \ref{matter_standard}, may be seen as \emph{``physical 
criteria set''} for a good monetary standard. In addition, the coefficients $c_n$ on equation \ref{eq:Conversion_E_to_M} can encapsulate the 
references to other sources of wealth, not directly associated with the ESC.

One might think that it is enough to build additionals power plants to ensure an artificial growth. Note that the standard is based on ESC, 
meanwhile, the product is still being offered in the conventional market and remains subject to trade rules. An excessive increase in supply 
would cause a drop in energy prices and next investments become less attractive. Note also that the inclusion of new power plants on the 
matrix involve years or decades of planning and construction and thus, the system exhibits a high level of self-protection against 
speculative attacks.

\section{Conclusions}\label{conclusions}

In this work, we did not intend to obtain a physical theory or to fit precisely the concepts and physical definitions with the 
macroeconomic dynamics or environment of finance. Also did not mean to explore in-depth the involved physical ideas. But in a simplified 
way, some physical characteristics of a monetary standard are identified. We also made a question about the process of value assigning to 
the material element of the standard that, analogously to the supply capacity idea, may be employed to the ESC-standard. 
Naturally, a change in focus of the conceptual economic evaluation has significant consequences throughout the economy. Nevertheless, 
the money circulation models and the inflation controlling are practical issues that remain open and need further studies to elucidate. 
However, it seems clear that the energy supply capacity, as explored here, is more interesting to a technological society than any millenary 
valuation conventions based on a specific kind of matter, and especially regarding the evaluation of long-term economic growth 
opportunities.

Given that the ESC-standard is based on a robust physical concept, it can be a possible alternative to a modern monetary standard sought by 
Nash (\citealp[p.7]{Nash2002}). Of course, an energy-based standard cannot be blindly adopted. There are important  political and 
geopolitical issues to consider, not discussed here. However, it can help to compensate for the shortage of gold and contribute to the 
establishment of more stabler currencies and an environment with lower volatility. In other hand, two factors will imply a significant 
change in the energy market: the automotive industry is signaling towards a strong introduction of electric-powered vehicles in the coming 
decades (\citealt{EV2017}), and the energetic market expansion upon renewable sources (\citealt{RES2017}). This scenario, coupled with the 
adoption of a monetary standard based on the energy supply capacity and with a definitive redemption to renewable energy power sources, may 
produce a promising environment for economic growth. It is necessary to be prepared for these further circumstances, even if it involves a 
small change of our current money concept.
\\

%\clearpage

% Bibliography.

\begin{doublespacing}   % Double-space the bibliography
\bibliographystyle{jf}
\bibliography{RHSbib}

\begin{thebibliography}{26}
\expandafter\ifx\csname natexlab\endcsname\relax\def\natexlab#1{#1}\fi

\bibitem[Argyle and Furnham(2013)]{PSYCHOLOGY}
Argyle, M., and A.~Furnham, 2013, {\em {The Psychology of Money}\/} (Taylor \&
  Francis).

\bibitem[Belke et~al.(2011)]{ENERGYCONSUMIPTION02}
Belke, A., et~al., 2011, {Energy consumption and economic growth: New insights
  into the cointegration relationship}, {\em Energy Economics\/} 33, 782--789.

\bibitem[{de A.L. Pereira} et~al.(2017)]{ECONOFISICA1}
{de A.L. Pereira}, E.~J., et~al., 2017, {Econophysics: Past and present}, {\em
  {Physica A: Stat.Mech. and its App.}\/} 473, 251--261.

\bibitem[Everett(2004)]{TIMETRAVEL02}
Everett, A., 2004, {Time travel paradoxes, path integrals, and the many worlds
  interpretation of quantum mechanics}, {\em Phys. Rev. D\/} 69, 124023.

\bibitem[Handa(2008)]{GOLDSTANDART}
Handa, J., 2008, {\em {Monetary Economics}\/} (Taylor {\&} Francis).

\bibitem[Hawking(1992)]{TIMETRAVEL01}
Hawking, S.~W., 1992, {Chronology protection conjecture}, {\em Phys. Rev. D\/}
  46, 603--611.

\bibitem[Hawking(1998)]{VIAGENS_TEMPO}
Hawking, S.~W., 1998, {\em {A Brief History of Time}\/}, Updated and expanded
  tenth anniversary edition (Bantam Books).

\bibitem[Hawking(2006)]{TEORIA_UNIFICADA}
Hawking, S.~W., 2006, {\em {The Theory of Everything: The Origin and Fate of
  the Universe}\/} (Phoenix Books).

\bibitem[Henriksen(2011)]{APLIC_RELATIVIDADE}
Henriksen, R.~N., 2011, {\em {Practical Relativity: From First Principles to
  the Theory of Gravity}\/} (Wiley).

\bibitem[IEA(2017{\natexlab{a}})]{EV2017}
IEA, 2017{\natexlab{a}}, {Global EV Outlook 2017}, {\em {IEA Publications}\/} .

\bibitem[IEA(2017{\natexlab{b}})]{RES2017}
IEA, 2017{\natexlab{b}}, {Market Report Series: Renewables 2017, Analysis and
  Forecasts to 2022}, {\em {IEA Publications}\/} .

\bibitem[Leff and Rex(2002)]{MAXWELL_DEMON}
Leff, H., and A.~F. Rex, 2002, {\em {Maxwell's Demon 2 Entropy, Classical and
  Quantum Information, Computing}\/} (CRC Press).

\bibitem[Nash(2002)]{Nash2002}
Nash, J.~F., 2002, Ideal money, {\em Southern Economic Journal\/} 69, 4--11.

\bibitem[{Nature Editorial}(2006)]{ECONOFISICA0}
{Nature Editorial}, 2006, {Econophysicists matter}, {\em Nature\/} 441,
  667--667.

\bibitem[{Nature Editorial}(2013)]{ECONOFISICA6}
{Nature Editorial}, 2013, {Net gains}, {\em Nature Physics\/} 9, 119--119.

\bibitem[Noether(1918)]{Noether1918}
Noether, E., 1918, Invariante variationsprobleme, {\em Nachrichten von der
  Gesellschaft der Wissenschaften zu Göttingen, Mathematisch-Physikalische
  Klasse\/} 1918, 235--257.

\bibitem[Ozturk et~al.(2010)]{ENERGYCONSUMIPTION01}
Ozturk, I., et~al., 2010, {Energy consumption and economic growth relationship:
  Evidence from panel data for low and middle income countries}, {\em Energy
  Policy\/} 38, 4422--4428.

\bibitem[Pokrovskii et~al.(2016)]{MONEYMODEL02}
Pokrovskii, V.~N., et~al., 2016, An elementary model of money circulation, {\em
  Physica A: Stat.Mech. and its App.\/} 463, 111--122.

\bibitem[Sardanashvily(2016)]{NOETHER}
Sardanashvily, G., 2016, {\em {Noether's Theorems: Applications in Mechanics
  and Field Theory}\/}, Atlantis Studies in Variational Geometry (Atlantis
  Press).

\bibitem[Schinckus(2013)]{ECONOFISICA2}
Schinckus, C., 2013, {Introduction to econophysics: towards a new step in the
  evolution of physical sciences}, {\em Contemporary Physics\/} 54, 17--32.

\bibitem[Schinckus et~al.(2018)]{MONEYMODEL01}
Schinckus, C., et~al., 2018, {Empirical justification of the elementary model
  of money circulation}, {\em Physica A: Stat.Mech. and its App.\/} 493,
  228--238.

\bibitem[Scott(1933)]{PROP_ENERGIA_ANTERIOR1}
Scott, H., 1933, {Technology smashes the price system}, {\em Harper's
  Magazine\/} 166, 129--144.

\bibitem[Sgouridis(2014)]{PROP_ENERGIA_ANTERIOR2}
Sgouridis, S., 2014, {Defusing the Energy Trap: The Potential of
  Energy-Denominated Currencies to Facilitate a Sustainable Energy Transition},
  {\em Frontiers in Energy Research\/} 2, 8.

\bibitem[Sgouridis and Kennedy(2010)]{PROP_ENERGIA_ANTERIOR3}
Sgouridis, S., and S.~Kennedy, 2010, Tangible and fungible energy: Hybrid
  energy market and currency system for total energy management. a masdar city
  case study, {\em Energy Policy\/} 38, 1749--1758, Energy Security - Concepts
  and Indicators with regular papers.

\bibitem[Urquhart(2016)]{BITCOIN01}
Urquhart, A., 2016, {The inefficiency of Bitcoin}, {\em Economics Letters\/}
  148, 80--82.

\bibitem[Yakovenko(2016)]{Yakovenko2016}
Yakovenko, V.~M., 2016, {Monetary economics from econophysics perspective},
  {\em Eur.Phys.J.Spec.Top.\/} 225, 3313--3335.

\end{thebibliography}
\end{doublespacing}

%\clearpage

% Print end notes
\renewcommand{\enotesize}{\normalsize}
% \begin{doublespacing}
%   \theendnotes
% \end{doublespacing}

\end{document}